\documentclass[showpacs,preprintnumbers,amsmath,amssymb,floatfix,superscriptaddress]{revtex4}  
\usepackage{graphicx}
\usepackage{dcolumn}
\usepackage{bm}
\usepackage{eepic,overpic}
\usepackage[dvips]{color}
\usepackage{times} 
\usepackage[normalem]{ulem}
\usepackage{lineno}
\usepackage[thinspace,thinqspace]{fancyunits}
\addunit{\eV}{eV}
\addunit{\EeV}{EeV}
\addunit{\yr}{yr}
\addunit{\VEM}{VEM}
\addunit{\inch}{inch}
\addunit{\pc}{pc}
\addunit{\Mpc}{Mpc}
\addunit{\events}{events}
\addunit{\Pascal}{Pa}
\begin{document}
\title{
  Observation of the suppression of the flux of cosmic rays above 
  \boldmath \unit{$4\times10^{19}$}{\eV}
}
\author{J.~Abraham}
\affiliation{Universidad Tecnol\'{o}gica Nacional, FR-Mendoza, 
Argentina}
\author{P.~Abreu}
\affiliation{LIP and Instituto Superior T\'{e}cnico, Lisboa, 
Portugal}
\author{M.~Aglietta}
\affiliation{Istituto di Fisica dello Spazio Interplanetario 
(INAF), Universit\`{a} di Torino and Sezione INFN, Torino, Italy}
\author{C.~Aguirre}
\affiliation{Universidad Catolica de Bolivia, La Paz, Bolivia}
\author{D.~Allard}
\affiliation{Laboratoire AstroParticule et Cosmologie, 
Universit\'{e} Paris 7, IN2P3/CNRS, Paris, France}
\author{I.~Allekotte}
\affiliation{Centro At\'{o}mico Bariloche, Comision Nacional de 
Energ\'{\i}a At\'{o}mica and Instituto Balseiro (CNEA-UNC), San Carlos 
de Bariloche, Argentina}
\author{J.~Allen}
\affiliation{New York University, New York, NY, USA}
\author{P.~Allison}
\affiliation{Ohio State University, Columbus, OH, USA}
\author{J.~Alvarez-Mu\~{n}iz}
\affiliation{Universidad de Santiago de Compostela, Spain}
\author{M.~Ambrosio}
\affiliation{Sezione INFN di Napoli, Napoli, Italy}
% ...
\collaboration{The Pierre Auger Collaboration}
\noaffiliation

\homepage{http://www.auger.org}
\date{\today}
\begin{abstract}
  The energy spectrum of cosmic rays above~\unit{$2.5\times10^{18}$}{\eV}, 
  derived from 20,000 events recorded at the Pierre Auger Observatory, 
  is described. The spectral index $\gamma$ of the flux, 
  $J\propto E^{-\gamma}$, at energies between \unit{$4\times10^{18}$}{\eV} 
  and \unit{$4\times10^{19}$}{\eV} is
  $2.69\pm0.02\usk\text{(stat)}\usk\pm0.06\usk\text{(syst)}$,
  steepening to
  $4.2\pm0.4\usk\text{(stat)}\usk\pm0.06\usk\text{(syst)}$ at higher
  energies, consistent with the prediction by Greisen and by Zatsepin
  and Kuz'min.
\end{abstract}
\maketitle 
We report a measurement of the energy spectrum of cosmic rays showing
that the flux is strongly suppressed above \unit{$4\times
  10^{19}$}{\eV}.  This is in accord with the 1966 prediction of
Greisen~\cite{bib:Greisen} and of Zatsepin and
Kuz'min~\cite{bib:Zatsepin} (GZK) that the spectrum should steepen
around \unit{$5\times 10^{19}$}{\eV} as cosmic rays from
cosmologically distant sources suffer energy losses when propagating
through the cosmic microwave radiation.  With an exposure twice that
of HiRes~\cite{bib:HiRes} and 4 times that of AGASA~\cite{bib:AGASA},
our evidence supports the recent report of the former.

The Pierre Auger Observatory, located near Malarg\"ue (Argentina) at
\unit{1400}{\meter} a.s.l., is used to measure the properties of
extensive air showers (EAS) produced by the highest-energy cosmic
rays.  At ground level the electrons, photons and muons of EAS can be
detected using instruments deployed in a large surface array.
Additionally, as EAS move through the atmosphere, ultra-violet light
is emitted from nitrogen excited by charged particles. This
\emph{fluorescence light} is proportional to the energy deposited by
the shower along its path~\cite{bib:edep1}.  The Observatory uses 1600
water-Cherenkov detectors, each containing 12 tonnes of water, viewed
by three $\unit{9}{"}$ photomultipliers, to detect the photons and
charged particles.  The surface detectors are laid out over
\unit{3,000}{\kilo\meter\Squared} on a triangular grid of
\unit{1.5}{\kilo\meter} spacing and is overlooked by 4 fluorescence
detectors. Each fluorescence detector (FD), located on the perimeter
of the area, houses 6 telescopes. EAS detected by both types of
detector are \emph{hybrid events} and play a key role in the analysis.
The field of view of each telescope is $30\degree$ in azimuth, and
$1.5\degree-30\degree$ in elevation.  Light is focused on a camera
containing 440 hexagonal pixels, of \unit{$18$}{\centi\meter\Squared},
at the focus of a \unit{11}{\meter\Squared} mirror.  The design and
status of the Observatory are described
in~\cite{bib:AugerNIM04,bib:ICRC07Dawson}.  Between 1 Jan 2004 to 31
Aug 2007 the numbers of telescopes increased from 6 to 24 and of
surface detectors from 154 to 1388. The analysis of data from this
period is described.

A cosmic ray of \unit{$10^{19}$}{\eV} arriving vertically typically
produces signals in 8 surface detectors. Using relative timing, the
direction of such an event is reconstructed with an angular accuracy
of about 1\degree{}~\cite{bib:ICRC07Ave}.  Signals are quantified in
terms of the response of a surface detector (SD) to a muon travelling
vertically and centrally through it (a \emph{vertical equivalent muon}
or VEM).  Calibration of each SD is carried out continuously with 2\%
accuracy~\cite{bib:bertouCalibration}.  The signals are fitted in each
event to find the VEM size at \unit{1000}{\meter},
$S(1000)$~\cite{bib:NewtonApPhys}.  The uncertainty in every $S(1000)$
is found, accounting for statistical fluctuations of the signals,
systematic uncertainties in the assumption of the fall-off of signal
with distance and the shower-to-shower
fluctuations~\cite{bib:ICRC07Ave}.  Above \unit{$10^{19}$}{\eV} the
uncertainty in $S(1000)$ is about 10\%.

The longitudinal development of EAS in the atmosphere is measured
using the fluorescence detectors. The light produced is detected as a
line of illuminated pixels in one or more FT cameras.  The positions
of these pixels and the arrival time of the light determine the shower
direction.  The signal, after correcting for attenuation due to
Rayleigh and aerosol scattering, is proportional to the number of
fluorescence photons emitted in the field of view of the pixel.
Cherenkov light produced at angles close to the shower axis can be
scattered towards the pixels: this contamination is accounted
for~\cite{bib:ICRC07UngerFDReco}.  A Gaisser-Hillas
function~\cite{bib:GaisserHillas} is used to reconstruct the shower
profile which provides a measurement of the energy of the EAS
deposited in the atmosphere.  To derive the primary energy, an
estimate of the missing energy carried into the ground by muons and
neutrinos must be made based on assumptions about the mass of cosmic
rays and of the appropriate hadronic model.  For a primary beam that
is a 50/50 mixture of protons and iron, simulations of showers with
the QGSJET01 model indicate a correction of
10\%~\cite{bib:Barbosa}. The systematic uncertainty is
4\%~\cite{bib:Pierog}.

Detailed understanding of the fluorescence emission is needed for
accurate energy determination. The absolute fluorescence yield in air
at \unit{293}{\kelvin} and \unit{1013}{\hecto\Pascal} from the
\unit{337}{\nano\meter} band is
\unit{$5.05\pm0.71$}{\text{photons}\per\mega\eV} of energy
deposited~\cite{bib:naganoFY}. The wavelength and pressure dependence
of the yield adopted follow~\cite{bib:airflyPressure}.  Systematic
uncertainties in the FD energy measurement have been
estimated. Measurements, made in combination with the fluorescence
detectors, are used to measure the quality and transmission properties
of the atmosphere. In particular, the vertical aerosol optical depth
(VAOD) profile~\cite{bib:ICRC07ben-zvi} is found every 15~min by
observing the light scattered from a centrally-located laser of an
energy equivalent to a few~\unit{$10^{19}$}{\eV} at
\unit{355}{\nano\meter}~\cite{bib:JINST06Fick} yielding an hourly
average. The average correction to $E_\text{FD}$ from the VAOD
measurement is +5\% at \unit{$3\times 10^{18}$}{\eV} rising to +18\%
at \unit{$5\times10^{19}$}{\eV}, reflecting the increase of the
average distance of such events from an FD.  The absolute calibration
of the telescopes is measured every few months and is constantly
adjusted using relative calibrations~\cite{bib:ICRC07Knapik}. The
largest uncertainties are in the absolute fluorescence yield (14\%),
the absolute calibration of the telescopes (10\%) and the
reconstruction method (10\%). Systematic uncertainties from
atmospheric aerosols, the dependence of the fluorescence spectrum on
temperature and on humidity are each at the 5\%
level~\cite{bib:ICRC07Dawson,bib:ICRC07Prouza}.  These uncertainties
are independent and added in quadrature give 22\% for $E_\text{FD}$.

The fluorescence detectors are operated on clear, moonless nights
limiting the duty cycle to 13\%. Showers detected by both the surface
array and the FD (hybrid events) are more precisely reconstructed than
surface array- or FD-only events~\cite{bib:ICRC07Dawson} and are
essential to the evaluation of systematic uncertainties.  The hybrid
events have an angular accuracy that improves from 0.8\degree{}
at~\unit{$3\times10^{18}$}{\eV} to 0.5\degree{}
above~\unit{$10^{19}$}{\eV}.  The surface array, with its near 100\%
duty cycle, gives the large sample used here.  The comparison of the
shower energy, measured using fluorescence, with the $S(1000)$ for a
subset of hybrid events is used to calibrate the energy scale for the
array.

\begin{figure}[!t]
\centering
  \begin{overpic}[width=0.46\textwidth]{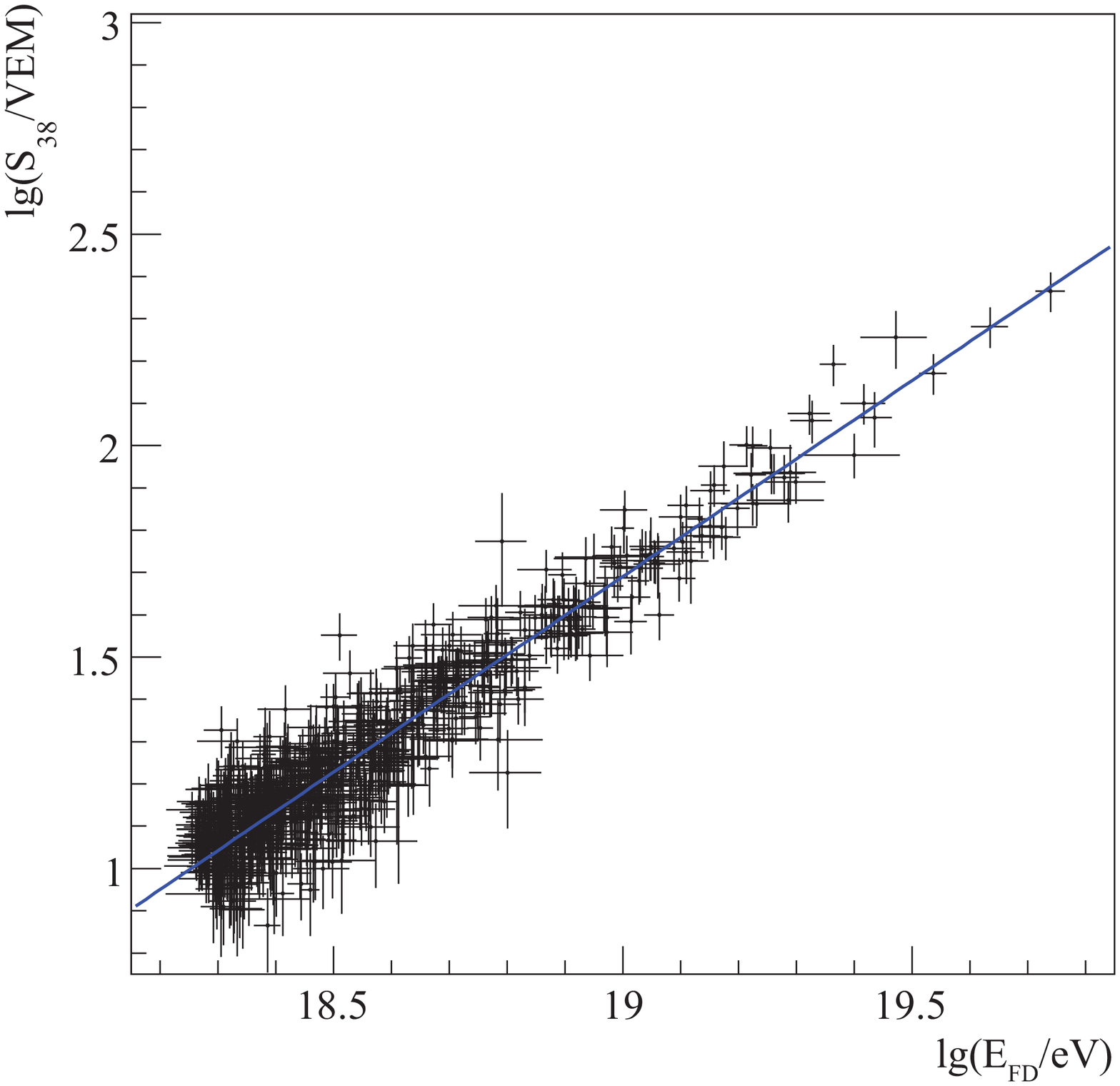}
    \put(15.8,52.5){\includegraphics[width=0.205\textwidth,clip=]{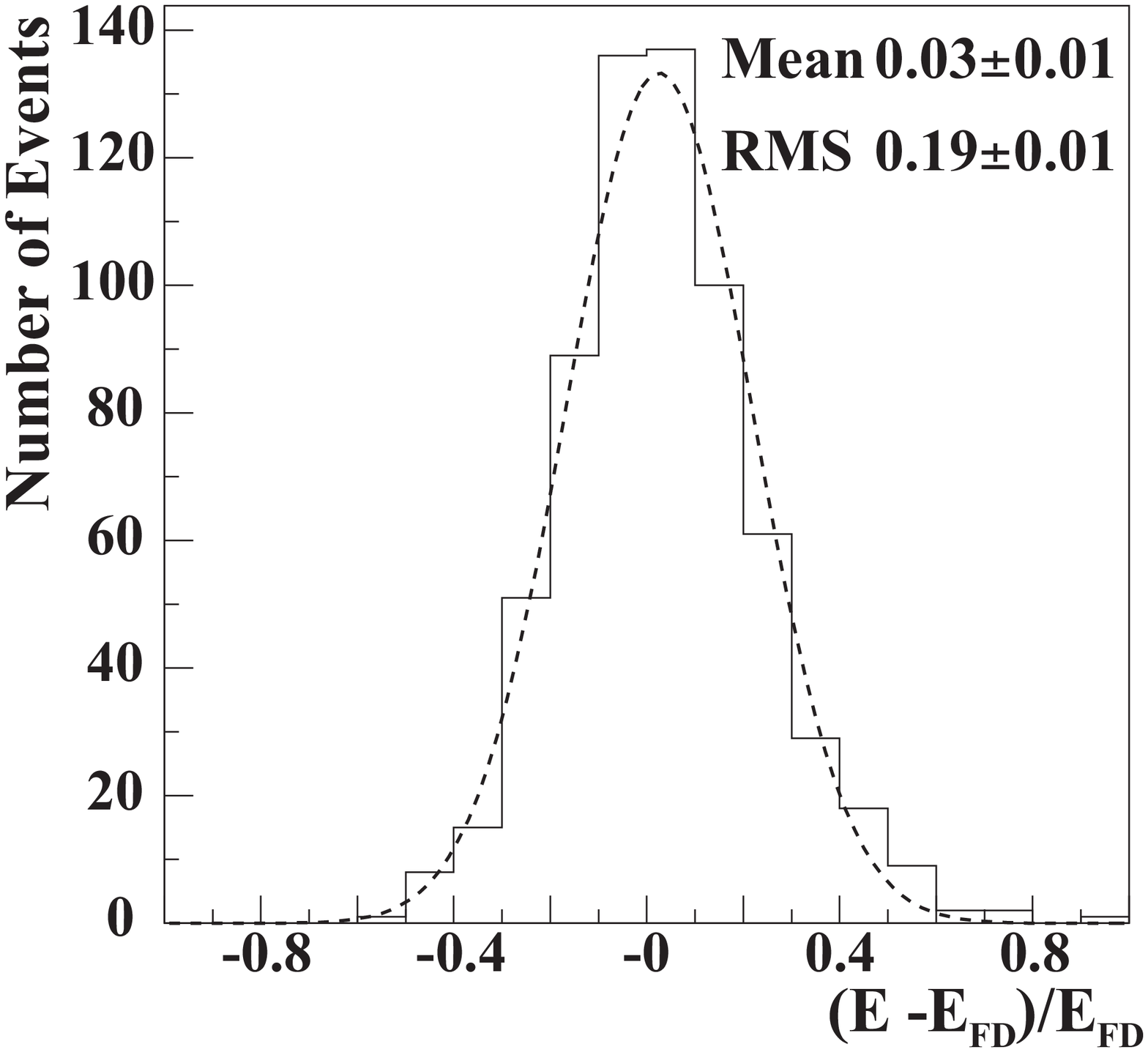}}
  \end{overpic}
  \caption{Correlation between $\lg S_{38^{\circ}}$ and $\lg
    E_\text{FD}$ for the 661 hybrid events used in the fit. The full
    line is the best fit to the data. The fractional differences
    between the two energy estimators are inset.
     \label{fig:SDCalibration}}
\end{figure}
Only events with zenith angles less than $60\degree$ are used
here. Candidate showers are selected on the basis of the topology and
time compatibility of the triggered
detectors~\cite{bib:ICRC05AllardTrigger}.  The SD with the highest
signal must be enclosed within an \emph{active hexagon}, in which all
six surrounding detectors were operational at the time of the
event. Thus it is guaranteed that the intersection of the axis of the
shower with the ground is within the array, and that the shower is
sampled sufficiently to make reliable measurements of $S(1000)$ and of
the shower axis. From the analysis of hybrid events, using only the
fall of the signal size with distance, these criteria result in a
combined trigger and reconstruction efficiency greater than 99\% for
energies above about \unit{$3\times10^{18}$}{\eV}; at
\unit{$2.5\times10^{18}$}{\eV} it is 90\%~\cite{bib:ICRC05Allard}.
The sensitive area has been calculated from the total area of the
hexagons active every second.

The decrease of $S(1000)$ with zenith angle arising from the
attenuation of the shower and from geometrical effects is quantified
by applying the \emph{constant integral intensity cut
  method}~\cite{bib:Hersil}, justified by the approximately isotropic
flux of primaries.  An energy estimator for each event, independent of
$\theta$, is $S_{38^\circ}$, the $S(1000)$ that EAS would have
produced had it arrived at the median zenith angle,
38\degree~\cite{bib:ICRC07Roth}.  Using information from the
fluorescence detectors the energy corresponding to each $S_{38^\circ}$
can be estimated almost entirely from data except for assumptions
about the missing energy. The energy calibration is obtained from a
subset of high-quality hybrid events, where the geometry of an event
is determined from the times recorded at an FD, supplemented by the
time at the SD with the highest signal, if it is within
\unit{750}{\meter} from the shower
axis~\cite{bib:ICRC07Unger,bib:ICRC07Perrone}.  It is also required
that a reduced $\chi^2$ is less than $2.5$ for the fit of the
longitudinal profile and that the depth of shower maximum be within
the field of view of the telescopes. The fraction of the signal
attributed to Cherenkov light must be less than 50\%.  Statistical
uncertainties in $S_{38^{\circ}}$ and $E_\text{FD}$ were assigned to
each event: averaged over the sample these were 16\% and 8\%,
respectively.

The correlation of $S_{38^\circ}$ with $E_\text{FD}$ is shown in
Fig.~\ref{fig:SDCalibration}, together with the least-squares fit of
the data to a power-law, $E_\text{FD} = a \cdot S_{38^{\circ}}^b$.
The best fit yields
\unit{$a=(1.49\pm0.06\text{\usk(stat)}\pm0.12\text{\usk(syst)})\times
  10^{17}$}{\eV} and $b=1.08\pm
0.01\text{\usk(stat)}\pm0.04\text{\usk(syst)}$ with a reduced $\chi^2$
of 1.1. $S_{38^\circ}$ grows approximately linearly with energy.  The
energy resolution, estimated from the fractional difference between
$E_\text{FD}$ and the derived SD energy, $E = a \cdot
S_{38^{\circ}}^b$, is shown inset. The root-mean-square deviation of
the distribution is 19\%, in good agreement with the quadratic sum of
the $S_{38^\circ}$ and $E_\text{FD}$ statistical uncertainties of
18\%.  The calibration accuracy at the highest energies is limited by
the number of events: the most energetic is
\unit{$\sim6\times10^{19}$}{\eV}.  The calibration at low energies
extends below the range of interest.

\begin{figure}[!t]
  \centering
  \includegraphics[width=0.445\textwidth]{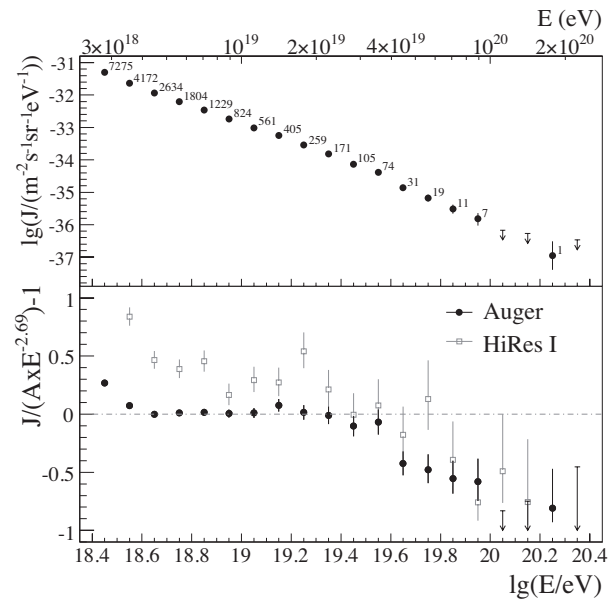}
  \caption{
    Upper panel:  The differential flux $J$ as a function of energy, 
    with statistical uncertainties.  Data are listed at ~\cite{bib:augerweb}.  
    Lower Panel: The fractional differences between Auger and HiRes I 
    data~\cite{bib:HiRes} compared with a spectrum with an index of 2.69.
    \label{fig:EnergySpectrum}}
\end{figure}

The energy spectrum based on $\sim$$20,000$ events is shown in
Fig.~\ref{fig:EnergySpectrum}.  Statistical uncertainties and 84\%
confidence-level limits are calculated according
to~\cite{bib:feldman}.  Systematic uncertainties on the energy scale
due to the calibration procedure are 7\% at \unit{$10^{19}$}{\eV} and
15\% at \unit{$10^{20}$}{\eV}, while a 22\% systematic uncertainty in
the absolute energy scale comes from the FD energy measurement.  The
possibility of a change in hadronic interactions or in the mean
primary mass above \unit{$6\times10^{19}$}{\eV} will be addressed with
more data.  In photon-initiated showers the value of $S(1000)$ is 2-3
times smaller than for nuclear primaries, so that a large photon flux
would change the spectrum. However, a limit to the photon-flux of 2\%
above \unit{$10^{19}$}{\eV} exists~\cite{bib:SDphotonLimit}.

The spectrum is fitted by a smooth transition function with the
suppression energy of \unit{$4\times10^{19}$}{\eV} defined as that at
which the flux falls below an extrapolated power law by 50\%.  To
examine the spectral shape at the highest energies, we fit a power-law
function between \unit{$4\times10^{18}$}{\eV} and \unit{$4\times
  10^{19}$}{\eV}, $J\propto E^{-\gamma}$, using a binned likelihood
method~\cite{bib:ICRC07Hague}. A power-law is a good parameterization:
the spectral index obtained is
$2.69\pm0.02\text{\usk(stat)}\pm0.06\text{\usk(syst)}$ (reduced
$\chi^2=1.2$), the systematic uncertainty coming from the calibration
curve. The numbers expected if this power-law were to hold above
\unit{$4\times10^{19}$}{\eV} or $10^{20}\eV$, would be $167\pm3$ and
$35\pm1$ while 69 events and 1 event are observed. The spectral index
above \unit{$4\times10^{19}$}{\eV} is
$4.2\pm0.4\usk\text{(stat)}\usk\pm0.06\usk\text{(syst)}$. A method
which is independent of the slope of the energy spectrum is used to
reject a single power-law hypothesis above
\unit{$4\times10^{18}$}{\eV} with a significance of more than 6
standard deviations~\cite{bib:ICRC07Hague}, a conclusion independent
of the systematic uncertainties currently associated with the energy
scale.

In Fig.~\ref{fig:EnergySpectrum} the fractional differences with
respect to an assumed flux $\propto E^{-2.69}$ are shown.  HiRes I
data~\cite{bib:HiRes} show a softer spectrum where our index is 2.69
while the position of suppression agrees within the quoted systematic
uncertainties. The AGASA data are not displayed as they are being
revised~\cite{bib:TeshimaRICAP}. The change of spectral index
indicated below \unit{$4\times10^{18}$}{\eV} will be discussed
elsewhere.

To summarize, we reject the hypothesis that the cosmic-ray spectrum
continues with a constant slope above \unit{$4\times10^{19}$}{\eV},
with a significance of 6 standard deviations. In a previous
paper~\cite{bib:Science}, we reported that sources of cosmic rays
above \unit{$5.7\times10^{19}$}{\eV} are extragalactic and lie within
75 Mpc. Taken together, the results suggest that the GZK prediction of
spectral steepening may have been verified. A full identification of
the reasons for the suppression will come from knowledge of the mass
spectrum in the highest-energy region and from reductions of the
systematic uncertainties in the energy scale which will allow the
derivation of a deconvolved spectrum.
\begin{acknowledgments}
We thank the technical and administrative staff in Malarg\"ue for
their exceptional dedication and the following organisations for
financial support:
Comisi\'on Nacional de Energ\'ia At\'omica, Fundaci\'on Antorchas,
Gobierno De La Provincia de Mendoza, Municipalidad de Malarg\"ue, NDM
Holdings and Valle Las Le\~nas, in gratitude for their continuing
cooperation over land access, Argentina; the Australian Research
Council; Conselho Nacional de Desenvolvimento Cient\'ifico e
Tecnol\'ogico (CNPq), Financiadora de Estudos e Projetos (FINEP),
Funda\c{c}\~ao de Amparo \`a Pesquisa do Estado de Rio de Janeiro
(FAPERJ), Funda\c{c}\~ao de Amparo \`a Pesquisa do Estado de S\~ao
Paulo (FAPESP), Minist\'erio de Ci\^{e}ncia e Tecnologia (MCT),
Brazil; AVCR AV0Z10100502 and AV0Z10100522, GAAV KJB300100801, GACR
202/06/P006, MSMT-CR LA08016, LC527 and 1M06002, Czech Republic;
Centre de Calcul IN2P3/CNRS, Centre National de la Recherche
Scientifique (CNRS), Conseil R\'egional Ile-de-France, D\'epartement
Physique Nucl\'eaire et Corpusculaire (PNC-IN2P3/CNRS), D\'epartement
Sciences de l'Univers (SDU-INSU/CNRS), France; Bundesministerium f\"ur
Bildung und Forschung (BMBF), Deutsche Forschungsgemeinschaft (DFG),
Finanzministerium Baden-W\"urttemberg, Helmholtz-Gemeinschaft
Deutscher Forschungszentren (HGF), Ministerium f\"ur Wissenschaft und
Forschung, Nordrhein-Westfalen, Ministerium f\"ur Wissenschaft,
Forschung und Kunst, Baden-W\"urttemberg, Germany; Istituto Nazionale
di Fisica Nucleare (INFN), Ministero dell'Istruzione,
dell'Universit\`a e della Ricerca (MIUR), Italy; Consejo Nacional de
Ciencia y Tecnolog\'ia (CONACYT), Mexico; Ministerie van Onderwijs,
Cultuur en Wetenschap, Nederlandse Organisatie voor Wetenschappelijk
Onderzoek (NWO), Stichting voor Fundamenteel Onderzoek der Materie
(FOM), Netherlands; Ministry of Science and Higher Education, Grant
Nos. 1 P03 D 014 30, N202 090 31/0623, and PAP/218/2006, Poland;
Funda\c{c}\~ao para a Ci\^{e}ncia e a Tecnologia, Portugal; Ministry
for Higher Education, Science, and Technology, Slovenian Research
Agency, Slovenia; Comunidad de Madrid, Consejer\'ia de Educaci\'on de
la Comunidad de Castilla La Mancha, FEDER funds, Ministerio de
Educaci\'on y Ciencia, Xunta de Galicia, Spain; Science and Technology
Facilities Council, United Kingdom; Department of Energy, Contract
No. DE-AC02-07CH11359, National Science Foundation, Grant No. 0450696,
The Grainger Foundation USA; ALFA-EC / HELEN, European Union 6th
Framework Program, Grant No. MEIF-CT-2005-025057, and UNESCO.

\end{acknowledgments}

\end{document}